\documentclass[pra,twocolumn,english,floatfix,superscriptaddress,nofootinbib]{revtex4-2}
\usepackage[english]{babel}
\usepackage{textcomp,xcolor,natbib}
\usepackage{dcolumn}
\usepackage{graphicx}
\graphicspath{{Figures/}}

\usepackage{amsmath, bbm}
\usepackage{latexsym}
\usepackage{amsfonts}   
\usepackage{amssymb}
\usepackage{array}      
\usepackage{epsfig}
\usepackage{txfonts}
\usepackage{xcolor,braket}
\usepackage[colorlinks=true,linkcolor=blue,urlcolor=blue,citecolor=blue]{hyperref}
\usepackage[normalem]{ulem}
\usepackage{soul}
\usepackage{dsfont}

\usepackage{xfrac}  
\usepackage{relsize}

\usepackage{amsthm}
\usepackage{mathrsfs, mathtools, amsmath}

\usepackage[normalem]{ulem}
\usepackage{cancel}
\usepackage{enumitem}

\usepackage{xspace}

\definecolor{darkgreen}{rgb}{0,0.5,0}

\newcommand{\Tr}{\operatorname{Tr}}

\begin{document}

\title{Multitime memory beyond the quantum regression theorem in sequential measurement statistics}
\author{Paolo Luppi}
\email{paolo.luppi@unimi.it}
\affiliation{Dipartimento di Fisica ``Aldo Pontremoli'', Universit\`a degli Studi di Milano, via~Celoria~16, I-20133 Milan, Italy}
\affiliation{INFN, Sezione di Milano, Via Celoria 16, I-20133 Milan, Italy}
\author{ Claudia Benedetti}
\email{claudia.benedetti@unimi.it}
\affiliation{Dipartimento di Fisica ``Aldo Pontremoli'', Universit\`a degli Studi di Milano, via~Celoria~16, I-20133 Milan, Italy}
\affiliation{INFN, Sezione di Milano, Via Celoria 16, I-20133 Milan, Italy}
\author{ Andrea Smirne }
\email{andrea.smirne@unimi.it}
\affiliation{Dipartimento di Fisica ``Aldo Pontremoli'', Universit\`a degli Studi di Milano, via~Celoria~16, I-20133 Milan, Italy}
\affiliation{INFN, Sezione di Milano, Via Celoria 16, I-20133 Milan, Italy}

\begin{abstract}
    We investigate the presence of memory in the sequential measurement statistics of an open quantum system,
    as witnessed by the departure from the quantum regression theorem (QRT), that is, the possibility to predict multitime probabilities
     from the one-time reduced dynamical map.
    For factorized initial states, we identify an exact decomposition of the two-time propagator into a QRT-like contribution, fully determined by the reduced dynamical map, and a memory term encoding system--environment correlations across the intervention; in the weak-coupling regime, the memory term yields an explicit second-order correction expressed in terms of the reduced map and bath correlation functions. Furthermore, we introduce an operational quantifier 
    of QRT violations based on the distance between exact and QRT-predicted joint probabilities.  Benchmarking the framework on a spin--boson model     
    and using a pseudomode embedding as nonperturbative reference, we comprehensively analyze the impact of spectral-density parameters, environmental temperature, and measurement protocols on the non-Markovianity of the multitime statistics.  
    Comparison with a one-time  
    quantifier shows that reduced-state non-Markovianity and 
    multitime memory are related but inequivalent: the latter, as probed through sequential statistics, is intrinsically protocol dependent and can become visible at higher temporal order even when two-time statistics remain compatible with QRT predictions.
\end{abstract}

\maketitle

\section{Introduction}
\label{sec:intro}
In many experimental settings involving open quantum systems, the quantities of interest are inherently multitime: they arise, for example, from control protocols, response and fluctuation functions, photon-counting statistics, 
and time-resolved spectroscopy \cite{Carmichael1993,Mukamel1995,Breuer2002,Gardiner2004,Rivas2011,Vacchini2024}. 
In these cases, the relevant information is contained in joint probability distributions or correlators associated with interventions performed at different times, and the reduced dynamical map, which fully characterizes the evolution of the system state at a single time, is in general not enough to determine such multitime quantities.

A standard construction to deal with multitime quantities is provided by the quantum regression theorem (QRT) \cite{Lax1963,Lax1968}, 
which expresses them by combining the same reduced dynamical map that describes one-time expectation values. This procedure is known to be valid under the weak-coupling and singular-coupling regime \cite{Dumcke1983} and, more generally, when system–environment correlations generated during the evolution do not affect the subsequent dynamics relevant for the considered quantities \cite{Swain1981}. 
However, finite bath correlation times, structured environments, strong coupling, and correlations built up prior to intermediate interventions can all lead to deviations from the QRT predictions, which calls for dedicated strategies to describe multitime objects in generic regimes~\cite{AlonsodeVega2005,IvanovBreuer2015,McCutcheon2016,Ban2018,Bonifacio2020,LonigroChruscinski202,Salamon2026}. 

Violations of the QRT can be associated to a notion of multitime non-Markovianity, in a sense close to the classical definition of non-Markovian stochastic processes, as the whole hierarchy of joint probability distributions cannot be reconstructed from one-time probabilities and propagators alone. This idea can be formalized by means of quantum stochastic processes \cite{Lindblad1979,Accardi1982} and process tensors \cite{Modi2012,Pollock2018b,Taranto2019,Milz2021,Giarmatzi2021}, where multitime statistics associated with sequences of interventions are treated as fundamental objects and Markovianity corresponds to a suitable factorization of the multitime correlations.
These multitime notions of memory are related to, but distinct \cite{Vacchini2011,Guarnieri2014,Milz2019b,LiHallWiseman2018} from, definitions of non-Markovianity based on the reduced dynamics. The latter are typically formulated in terms of properties of the dynamical map, such as P- or CP-divisibility \cite{WolfCirac2008,RivasHuelgaPlenio2010,Wissmann2015,Breuer2016,Chruscinski2022}, and characterize memory effects at the level of the system states. By contrast, QRT violations and, more generally, quantum-stochastic-processes and process-tensor approaches are defined in terms of sequential measurements and joint outcome distributions, so that they provide an operational notion of memory that 
is not fully determined by one-time properties of the reduced evolution.
Recent studies of spin-boson dynamics have shown that single-intervention process tensors can diagnose quantum memory more informatively than reduced-map criteria alone \cite{BackerLinkStrunz2025}, while a complementary line of work relates violations of QRT in two-point correlations to quantum features in the non-Markovian evolution, including system-environment entanglement \cite{SantosGuhneNimmrichter}.

In this work, we investigate multitime memory, as witnessed by the breakdown of QRT in sequential measurement statistics, 
for fixed projective measurements. First, using projection-operator techniques adapted to dynamics with intermediate interventions \cite{IvanovBreuer2015}, we derive an exact decomposition of the two-time propagator into a QRT-like term, fully determined by the reduced dynamical map, and a complementary contribution that fully encodes deviations from the QRT into system–environment correlations across the intervention. In the weak-coupling regime, this memory contribution can be expanded perturbatively, leading to an explicit second-order correction expressed in terms of the reduced map and bath correlation functions. On the operational side, we formulate the problem in terms of quantum instruments and introduce a quantifier of QRT breakdown based on the Kolmogorov distance between exact and QRT joint probabilities.

As a benchmark, we apply this framework to the spin–boson model with a Lorentzian spectral density, using the pseudomode embedding as a nonperturbative reference \cite{Garraway1997,Tamascelli2018}. This allows us to assess the accuracy of the perturbative correction and to analyze the dependence of QRT deviations on time, model parameters, temperature, and measurement protocols. We further compare the resulting multitime characterization with a one-time witness of P-divisibility reconstructed from the same reduced dynamics. The comparison shows that one-time and multitime notions capture different aspects of memory in the evolution: while they are both enhanced by a stronger coupling leading to large system-environment correlations, multitime memory depends on the measurement protocol and may become visible at higher temporal order even when lower-order quantities remain close to the corresponding Markovian predictions.

The remainder of the paper is organized as follows. In Sec.~\ref{sec:oqs} 
we introduce the open quantum system framework and the description of multitime objects therein, corresponding to correlators
and sequential-measurement joint probability distributions. 
In Sec.~\ref{sec:analytic} we present an exact decomposition of the two-time propagator into a QRT contribution and a correction, 
which we explicitly work out at leading order in the coupling strength. 
In Sec.~\ref{sec:numerics} we apply our results to the paradigmatic spin--boson model, quantifying the breakdown of the QRT description,
across different dynamical regimes. In particular, we compare QRT violations with a one-time divisibility-based witness of non-Markovianity, and analyze protocol dependence, temperature effects, and higher-order temporal statistics. 
Section~\ref{sec:conclusions} contains the conclusions and future perspectives of our work.

\section{Open quantum dynamics}
\label{sec:oqs}

\subsection{System--bath dynamics}\label{sec:sbd}
We consider a system $S$ with Hilbert space $\mathcal{H}_S$ coupled to an environment (bath) $B$ with Hilbert space $\mathcal{H}_B$, such that the  total space is $\mathcal{H}=\mathcal{H}_S\otimes\mathcal{H}_B$.
The joint system-environment state $\rho(t)$ evolves according to the Liouville--von Neumann equation \cite{Breuer2002}
\begin{equation}
\frac{d}{dt}\rho(t)=\mathcal{L}[\rho(t)],\qquad
\mathcal{L}=\mathcal{L}_S+\mathcal{L}_B+\mathcal{L}_{SB},
\label{eq:Liouvillian_total}
\end{equation}
where $\mathcal{L}_X[\rho]=-i[H_X,\rho]$ for $X\in\{S,B,SB\}$
and $H_S$ ($H_B$) describes the free system (bath) Hamiltonian, while $H_{SB}$ describes the interaction term (units $\hbar=1$);
indeed, $\mathcal L$ denotes the Liouvillian on the global $S-B$ system. We also use the free Liouvillian $\mathcal{L}_0:=\mathcal{L}_S+\mathcal{L}_B$ and the interaction-picture coupling superoperator
\begin{equation}
\mathcal{L}_{I}(t):=e^{-t\mathcal{L}_0}\,\mathcal{L}_{SB}\,e^{t\mathcal{L}_0}.
\label{eq:LI_def}
\end{equation}
Throughout we assume a coupling that can be written in the form
\begin{equation}
H_{SB}=\lambda\sum_a S_a\otimes B_a,
\label{eq:HSB_bilinear}
\end{equation}
where $\lambda$ is the coupling-strength parameter.
We fix a bath reference state $R$ satisfying
\begin{equation}
\mathcal{L}_B R=0,
\label{eq:R_stationary}
\end{equation}
i.e., $R$ is stationary under the free bath dynamics.
Without loss of generality, we take the bath interaction operators to have vanishing mean in $R$,
\begin{equation}
\mathrm{Tr}_B[B_a R]=0 \qquad \forall a  ,
\label{eq:Ba_zero_mean}
\end{equation}
since any nonzero mean can be absorbed into a renormalization of $H_S$ \cite{Rivas2011}.
Further assuming that the bath is Gaussian, its influence on the open-system dynamics is fully enclosed in the correlation functions
\begin{equation}
C_{ab}(t):=\mathrm{Tr}_B\!\left[B_a(t)\,B_b\,R\right], \quad
    B_a(t):=e^{-t\mathcal{L}_B}[B_a],
    \label{eq:bath_corr_def}
\end{equation}
which are referred to the environmental interaction operators evolving under the free bath dynamics.

The reduced state associated with the open system is obtained via the partial trace, $\rho_S(t)=\Tr_B[\rho(t)]$.
In the following we will consider factorized initial states 
of the form 
\begin{equation}\label{eq:in}
\rho(0)=\rho_S(0)\otimes R,
\end{equation}
so that we can define the
(one-time) reduced dynamical map $\Lambda_S(t)$, $t\geq0$, by
\begin{equation}
\rho_S(t)=\Lambda_S(t)[\rho_S(0)]
:=\Tr_B\!\left[e^{t\mathcal{L}}\big[\rho_S(0)\otimes R\big]\right],
\label{eq:Phi_def}
\end{equation}
which is a completely positive and trace-preserving (CPTP) map for each fixed $t$.

Standard definitions of one-time Markovianity are related to the notion of \emph{divisibility}. 
Assume that, for all $t_2\ge t_1\ge 0$, there exists an intermediate map $\mathcal{V}(t_2,t_1)$, i.e., a propagator, such that
\begin{equation}
\Lambda_S(t_2)=\mathcal{V}(t_2,t_1)\circ \Lambda_S(t_1);
\label{eq:divisibility_def}
\end{equation}
indeed, if $\Lambda_S(t_1)$ is invertible, the propagator can be identified with
\begin{equation}
\mathcal{V}(t_2,t_1)=\Lambda_S(t_2)\circ \Lambda_S(t_1)^{-1}.
\label{eq:intermediate_map_invertible}
\end{equation}
If each $\mathcal{V}(t_2,t_1)$ can be chosen positive and trace-preserving, the dynamics is \emph{P-divisible} \cite{Vacchini2011,Wissmann2015,Breuer2016};
if, in addition, each $\mathcal{V}(t_2,t_1)$ is completely positive, the dynamics is \emph{CP-divisible} \cite{RivasHuelgaPlenio2010}.
Both notions have been used to define one-time quantum (non)-Markovianity; in the following, we will focus in particular
on the investigation of P-divisibility.
This choice is motivated by the comparison with multitime Markovianity: throughout the paper, the latter is assessed via sequential
measurement statistics generated by instruments acting on the system alone, and
P-divisibility provides a weaker requirement directly testable at the level of system states, whereas
CP-divisibility is a stronger condition
whose operational meaning involves the presence of an ancilla \cite{Chruscinski2011}.

\subsection{Two-time correlators and sequential probabilities}
\label{subsec:twotime_objects}
We now move to the multitime scenario and introduce the notion of Markovianity
we will use throughout our analysis, focusing on sequential protocols where an intervention at $t_1$ is followed
by a readout at $t_2$,
for $0\le t_1\le t_2$. 
\\

Given an intervention $\mathcal{A}_1$ at time $t_1$ on the open system,
we define the \emph{conditioned two-time propagator}, for $0\le t_1\le t_2$,
\begin{align}
\Phi_{\mathcal{A}_1}(t_2,t_1)[\rho]
&:= \mathrm{Tr}_B\!\big[e^{(t_2-t_1)\mathcal{L}}\,
\circ\left(\mathcal{A}_1\otimes\mathcal{I}_B\right)\circ\, e^{t_1\mathcal{L}}\left[\rho\otimes R\right]\big],
\label{eq:PhiA_def}
\end{align}
where here and in the following we denote as $\mathcal{I}$ the identity map, 
and we used the total propagator $e^{t\mathcal{L}}$ defined by Eq.(\ref{eq:Liouvillian_total}).
Taking into account also the readout at time $t_2$, we can further define the two-time quantity 
\begin{equation}
F_{\mathcal{A}_2,\mathcal{A}_1}(t_2,t_1)
:= \Tr_S\Big[\mathcal{A}_2\circ\Phi_{\mathcal{A}_1}(t_2,t_1)[\rho_S(0)]\Big],
\label{eq:C_A_def}
\end{equation}
which covers the two main cases of interest for our investigation.
Choosing $\mathcal{A}_i =L_{A_i}$ the left multiplication by $A_i$, i.e.,
$L_{A_i}[\rho]=A_i \rho$,
Eq.~\eqref{eq:C_A_def} yields the time-ordered correlators 
\begin{equation} \label{eq:a2a1}
\langle A_2(t_2)A_1(t_1)\rangle = F_{A_2,A_1}(t_2,t_1)
= \Tr_S\!\Big[A_2 \Phi_{A_1}(t_2,t_1)[\rho_S(0)] \Big],
\end{equation}
i.e., the two-time correlators associated with the operators $A_1$ and $A_2$, which define, for example, linear-response
functions and emission spectra \cite{Mukamel1995,Kubo1966,Clerk2010}.

Instead, let $\{\mathcal{M}^{(1)}_{a_1}\}_{a_1 \in O_1}$ and $\{\mathcal{M}^{(2)}_{a_2}\}_{a_2 \in O_2}$ be instruments at $t_1$ and $t_2$, i.e.
collections of \ CP
trace-nonincreasing maps such that $\sum_{a_1}\mathcal{M}^{(1)}_{a_1}$ and $\sum_{a_2}\mathcal{M}^{(2)}_{a_2}$ are CPTP,
and that describe the statistics of a two-time sequential measurement with outcomes $(a_1,a_2) \in O_1 \times O_2$. 
The joint probability for outcomes $(a_1,a_2)$ at $(t_1,t_2)$ is obtained by choosing
$\mathcal{A}_i=\mathcal{M}^{(i)}_{a_i}$ and applying the trace after the second CP map,
\begin{align}\label{eq:seq_prob_def_ij}
p(a_2,t_2;\,a_1,t_1)
&=  F_{\mathcal{M}^{(2)}_{a_2},\mathcal{M}^{(1)}_{a_1}}(t_2,t_1) \\
&=\Tr_S\!\left[
\mathcal{M}^{(2)}_{a_2}\circ \Phi_{\mathcal{M}^{(1)}_{a_1}}(t_2,t_1)[\rho_S(0)]
\right].\nonumber
\end{align}
From this probability, we can obtain all the information about the statistics
of the sequential measurement; for example, the two-time expectation value is given by
\begin{equation}
\overline{A_2(t_2)A_1(t_1)}
:=
\sum_{a_1,a_2} a_1\,a_2\,p(a_2,t_2;\,a_1,t_1).
\label{eq:seq_moment_def_ij}
\end{equation}

We stress that even when the same quantities $A_1$ and $A_2$ are considered, correlators and
sequential measurement statistics generally differ because the latter include an explicit state transformation
at the intermediate time, whereas the correlator does not.
Explicitly, let us consider the two self-adjoint operators $A_1=A^{\dagger}_1$ and $A_2=A^{\dagger}_2$, with spectral decompositions
\begin{equation}
A_i=\sum_{a_i} a_{i} \Pi_{a_i}^{(i)}, \quad i=1,2,
\end{equation}
with $\Pi_{a_i}^{(i)}$ the projector into the eigenspace of the eigenvalue $a_i$ of the operator $A_i$, 
along with the projective measurement associated with it via the L\"uders instrument \cite{Heinosaari2011},
\begin{equation}
   \mathcal{M}^{(i)}_{a_i}[\rho]:=\Pi_{a_i}^{(i)} \rho \Pi_{a_i}^{(i)}.
\end{equation}
Introducing the full-dephasing map associated with the first intermediate measurement,
\begin{equation}
  \mathcal{D}_{A_1}[\rho]:=\sum_{a_1}\mathcal{M}^{(1)}_{a_1}[\rho] = \sum_{a_1} \Pi_{a_1}^{(1)}\rho\Pi_{a_1}^{(1)},
\end{equation}
we can express
the discrepancy between the correlator and the sequential-measurement average
as
\begin{eqnarray}
&&\langle A_2(t_2)A_1(t_1)\rangle - \overline{A_2(t_2)A_1(t_1)} \\
&&=\Tr_S\!\left[A_2\,\Phi_{(\mathcal{I}-\mathcal{D}_{A_1}) \circ  L_{A_1}}(t_2,t_1)[\rho_S(0)]\right].\nonumber
\end{eqnarray}
This relation shows
that the mismatch is entirely due to the off-diagonal (coherence) component
$(\mathcal{I}-\mathcal{D}_{A_1}) \left[A_1\rho_S(t_1)\right]$ removed by the L\"uders update, but potentially converted into
populations relevant to $A_2$ by the subsequent dynamics.
The sequential statistics depends on the choice of the instrument, and a full agreement with the correlators can occur 
only for non-invasive interventions or when the coherences in the measurement basis at $t_1$ vanish or are dynamically irrelevant for the readout measurement at time $t_2$
\cite{Smirne2019}.

The Markovianity criterion associated with multitime objects we consider consists in the validity of the quantum regression theorem (QRT)
\cite{Lax1968,Breuer2002}. In our notation and restricting to two-time quantities, the latter means that one can perform the replacement
\begin{equation}
\Phi_{\mathcal{A}_1}(t_2,t_1) \mapsto 
\Lambda_S(t_2-t_1)\circ \mathcal{A}_1\circ \Lambda_S(t_1),
\label{eq:QRT_like_factorization}
\end{equation}
into $F_{\mathcal{A}_2,\mathcal{A}_1}(t_2,t_1)$ defined in Eq.(\ref{eq:C_A_def}), for any initial state $\rho_S(0)$
and readout map $\mathcal{A}_2$. Microscopically, such a replacement is justified when the system-environment correlations built up by the interaction up to time $t_1$ are irrelevant for the quantities at time $t_2$ \cite{Swain1981}.
Indeed, the difference between averages of sequential measurements and two-time correlators manifests itself in the applicability of the QRT, 
as there are situations where it can be applied to the latter but not to the former, even when the same quantities are involved. 
In particular, for a two-level system undergoing pure dephasing due to the interaction with a single continuous degree of freedom 
initially distributed according to a Lorentzian profile, the QRT holds for any two-time correlator \cite{Guarnieri2014},
while it fails for the average of sequential projective measurements performed in a basis different from the dephasing one \cite{Smirne2019,Milz2020}.

\section{Beyond-QRT two-time propagators}
\label{sec:analytic}
\subsection{Exact expression via projection superoperators}
\label{subsec:analytic_decomposition}
The expression of the two-time expectation values can be rewritten by means of the projection superoperators,
to single out in full generality the term leading to a violation of the QRT.\\
In particular, introducing
the standard projection superoperators \cite{Breuer2002}
\begin{equation}
\mathcal{P}[\rho] := \Tr_B[\rho]\otimes R,\qquad 
\mathcal{Q}:=\mathcal{I}-\mathcal{P}
\label{eq:PQ_def_here}
\end{equation}
that act on operators on $S{+}B$,
and replacing the identity  $\mathcal I=\mathcal P+\mathcal Q$ in the definition of 
the conditioned two-time propagator Eq.(\ref{eq:PhiA_def}),
we can rewrite it 
as:
\begin{align}
\Phi_{\mathcal A_1}(t_2,t_1)[\rho]
&=
\Tr_B\!\big[e^{(t_2-t_1)\mathcal{L}}\circ
(\mathcal{A}_1\otimes\mathcal{I}_B)\,\nonumber\\
&\quad\circ(\mathcal P+\mathcal Q)\circ
e^{t_1\mathcal{L}}\left[\rho\otimes R\right]\big] \nonumber
\\
&=: \Phi^{(P)}_{\mathcal A_1}(t_2,t_1)[\rho]\;+\;\Phi^{(Q)}_{\mathcal A_1}(t_2,t_1)[\rho].\label{eq:exact_PQ_split}
\end{align}
Since
\begin{equation}
\mathcal P e^{t_1\mathcal L}\big[\rho\otimes R\big]=\Lambda_S(t_1)[\rho]\otimes R
\end{equation}
by definition of $\Lambda_S(t)$, we get
\begin{align}
\Phi^{(P)}_{\mathcal A_1}(t_2,t_1)[\rho]
&=
\Tr_B\!\big[e^{(t_2-t_1)\mathcal{L}}\,
\circ(\mathcal{A}_1\otimes\mathcal{I}_B)\,
\left[\Lambda_S(t_1)[\rho]\otimes R\right]\big]
\nonumber\\
&=
\Lambda_S(t_2-t_1)\circ\mathcal A_1\circ\Lambda_S(t_1)[\rho],
\label{eq:exact_QRT_like_part}
\end{align}
i.e., $\Phi^{(P)}_{\mathcal A_1}(t_2,t_1)$
exactly corresponds to a term in the form given by the QRT -- see Eq.(\ref{eq:QRT_like_factorization}).
Therefore,
\begin{align}
\Phi^{(Q)}_{\mathcal A_1}(t_2,t_1)[\rho]
&:=
\Tr_B\!\big[e^{(t_2-t_1)\mathcal{L}}\circ
(\mathcal{A}_1\otimes\mathcal{I}_B)\circ
\mathcal Q\circ
e^{t_1\mathcal{L}}\big[\rho\otimes R\big]\big]\label{eq:exact_memory_part}
\end{align}
is the QRT-violating contribution: it isolates the part of the two-time propagator that is not fixed by the
reduced state and the dynamical map only. 
 The size of the possible QRT violation is therefore controlled jointly by (i) how many
system-environment correlations and changes
in the environmental state are present at time $t_1$ -- i.e., 
the magnitude of $\mathcal Q\rho_{SB}(t_1)$ -- and (ii) how strongly the chosen
intervention map can convert them into an observable effect for the chosen readout at time $t_2$. 
In particular, the back-action of the first measurement and the readout share a nontrivial interplay with the correlation built up by the dynamics: an intervention may amplify QRT deviations by
conditioning on bath-correlated degrees of freedom, or suppress them when it reduces precisely those
coherences that would feed into the later readout.

The term $\Phi^{(Q)}_{\mathcal A_1}$ admits a double-convolution representation
\begin{align}
\Phi^{(Q)}_{\mathcal A_1}(t_2,t_1)
&=
\int_{0}^{t_2-t_1}\! d\tau_2 \int_{0}^{t_1}\! d\tau_1\;
\Lambda_S(t_2-t_1-\tau_2) \nonumber\\
&\circ \mathcal K^{\mathcal A_1}(\tau_2,\tau_1)\circ \Lambda_S(t_1-\tau_1),
\label{eq:main_double_convolution}
\end{align}
for the multitime kernel (here and in the following we imply the composition $\circ$ within the kernel)
\begin{align}
\mathcal K^{\mathcal A_1}(\tau_2,\tau_1)[\rho]
:=&
\Tr_B\!\big[
\mathcal L_{SB}\,e^{\tau_2\mathcal Q\mathcal L}\,\mathcal Q (\mathcal{A}_1\otimes\mathcal{I}_B)\nonumber\\
&
 e^{\tau_1 \mathcal Q\mathcal L}\,
 \mathcal Q \mathcal L_{SB}\,[\rho\otimes R]
\big],
\label{eq:app_kernel_LSB_form}
\end{align}
which can be derived by applying the
projection-operator multitime-kernel framework of Ref.~\cite{IvanovBreuer2015}, developed there for
multitime correlation functions. The same technique can be employed identically for a two-time
propagator with a generic intermediate intervention $\mathcal A_1$, 
which makes the connection to QRT violations of multitime probabilities explicit.

In fact, focusing on the case where the interventions consist of sequential measurements, 
we quantify the difference between the exact joint probability distribution $p$ 
given by Eq.(\ref{eq:seq_prob_def_ij}) 
and the joint distribution $p_{\rm QRT}$ obtained by applying the QRT -- i.e., 
replacing Eq.(\ref{eq:QRT_like_factorization}) into Eq.(\ref{eq:seq_prob_def_ij}) -- by means of
the Kolmogorov distance,
\begin{align}\label{eq:TV_def0}
&\varepsilon_{\rm QRT}(t_2,t_1):=
\frac{1}{2}\sum_{a_1,a_2}\big|\,p(a_2,t_2;\,a_1,t_1)-p_{\rm QRT}(a_2,t_2;\,a_1,t_1)\,\big|. 
\end{align}
Note that
since $p$ and $p_{\rm QRT}$ are normalized joint distributions, $\varepsilon_{\rm QRT}(t_2,t_1)\in[0,1]$,
with $\varepsilon_{\rm QRT}=0$ if and only if they coincide. 
Using Eqs.(\ref{eq:exact_PQ_split}) and (\ref{eq:exact_QRT_like_part}), such a difference can be compactly expressed as
\begin{align}\label{eq:TV_def}
&\varepsilon_{\rm QRT}(t_2,t_1) = 
\frac{1}{2}\sum_{a_1,a_2}\left|\Tr_S\!\left[
\mathcal{M}^{(2)}_{a_2}\circ \Phi^{(Q)}_{\mathcal{M}^{(1)}_{a_1}}(t_2,t_1)[\rho_S(0)]
\right]\right|,
\end{align}
which directly relates the deviations from the QRT expression in the joint probability distribution to microscopic details
of the system-environment interaction.

In the following analysis, we will also consider the $t_1$-average QRT violation, for a fixed final time $t_2=t_f$, namely,
\begin{equation}
\bar{\varepsilon}_{\rm QRT}(t_f)
:=
\frac{1}{t_f}\int_{0}^{t_f}\varepsilon_{\rm QRT}(t_f,t_1)\,dt_1.
\label{eq:epsQRTavg}
\end{equation}
Operationally, $\bar{\varepsilon}_{\rm QRT}(t_f)$ is the expected QRT error when the first intervention time $t_1$
is drawn uniformly at random in $[0,t_f]$ and it is indeed more suitable for a comparison with the single-time non-Markovianity.

\subsection{Second-order expression}\label{sec:soe}
The explicit evaluation of the two-time quantity in Eq.(\ref{eq:C_A_def}) is in general as complicated
as the solution of the global dynamics. In the next Section, 
we will perform the numerical simulation, in generic parameter regimes,
for the spin-boson model we are interested in. Here, instead, we report the lowest order of a perturbative expansion
that can be applied to any model in the weak coupling regime.

Since $\mathcal L_{SB}=O(\lambda)$, the leading order of the kernel is $O(\lambda^2)$, which is obtained by replacing 
the $\mathcal Q$-projected
propagators by their zeroth-order approximation in $\lambda$,
\begin{equation}
e^{\mathcal Q(\mathcal L_0+\mathcal L_{SB})\,\tau}
=
e^{\mathcal Q\mathcal L_0\tau}+O(\lambda),
\label{eq:Born_truncation_propagator}
\end{equation}
so that 
\begin{align}\label{eq:app_kernel_second}
&\mathcal K^{\mathcal A_1}(\tau_2,\tau_1)[\rho]
:=
\Tr_B\!\big[
\mathcal L_{SB}\,e^{\mathcal Q\mathcal L_0\,\tau_2}\,
\mathcal Q (\mathcal{A}_1\otimes\mathcal{I}_B)\\
&e^{\mathcal Q\mathcal L_0\,\tau_1}\,
\mathcal Q\mathcal L_{SB}\,[\rho\otimes R]
\big] + O(\lambda^3) =: \tilde{\mathcal{K}}^{\mathcal A_1}(\tau_2,\tau_1)[\rho] + O(\lambda^3).
\nonumber
\end{align}
To proceed further, we note that
\begin{equation}
[\mathcal L_0,\,\mathcal Q] = 0,
\label{eq:app_L0_Q_commute}
\end{equation}
since, for any $S-B$ operator $X$,
\begin{align}
&\mathcal P\mathcal L_0[X]
= \Tr_B[\mathcal L_0 X]\otimes R = \big(\mathcal L_S[\Tr_B X]
  + \Tr_B[\mathcal L_B X]\big)\otimes R \nonumber\\
&= \mathcal L_S[\Tr_B X]\otimes R
= \mathcal L_0[\Tr_B[X]\otimes R]
= \mathcal L_0\mathcal P[X], \nonumber
\end{align}
where we used that
$\Tr_B[\mathcal L_B X] = 0$
(cyclicity of $\Tr_B$) and $\mathcal L_B R = 0$. 
Exploiting the idempotency relation $\mathcal Q^2 = \mathcal Q$ and the power series of the exponential, 
we then have
\begin{equation}
e^{\mathcal Q\mathcal L_0\,\tau}\,\mathcal Q X
= e^{\mathcal L_0\tau}\,\mathcal Q X.
\label{eq:app_free_on_Q}
\end{equation}
Moreover, Eq.(\ref{eq:Ba_zero_mean}) implies
\begin{equation}
\mathcal P\mathcal L_{SB}[\rho\otimes R]
= -i\lambda\sum_a [S_a,\rho]\,\Tr_B[B_a R]\;\otimes\; R
= 0,
\end{equation}
while, using once more Eq.(\ref{eq:app_L0_Q_commute}) and $\mathcal P (\mathcal{A}_1\otimes\mathcal{I}_B) = (\mathcal{A}_1\otimes\mathcal{I}_B) \mathcal P$, one has
\begin{equation}\label{eq:aux1}
   \mathcal P (\mathcal{A}_1\otimes\mathcal{I}_B) e^{\mathcal L_0 \tau_1} \mathcal{Q} [X] = 0. 
\end{equation}

All in all, replacing Eqs.(\ref{eq:app_free_on_Q})-(\ref{eq:aux1}) (along with $\mathcal Q = \mathcal I - \mathcal P$) 
into Eq.(\ref{eq:app_kernel_second}), we arrive at the following expression of the kernel's leading order:
\begin{align}
\tilde{\mathcal{K}}^{\mathcal A_1}(\tau_2,\tau_1)[\rho]
=&
\Tr_B\!\big[
\mathcal L_{SB}\,e^{ \mathcal L_0\,\tau_2}\,
(\mathcal{A}_1\otimes\mathcal{I}_B)e^{\mathcal L_0\,\tau_1}\,
\mathcal L_{SB}\,(\rho\otimes R)
\big],
\label{eq:Born_dressed_kernel_def}
\end{align}
and therefore at the following leading expression of the QRT-violating term:
\begin{align}
\tilde{\Phi}^{(Q)}_{\mathcal A_1}(t_2,t_1)
&=
\int_{0}^{t_2-t_1}\! d\tau_2 \int_{0}^{t_1}\! d\tau_1\;
\Lambda_S(t_2-t_1-\tau_2) \nonumber\\
&\circ \tilde{\mathcal K}^{\mathcal A_1}(\tau_2,\tau_1)\circ \Lambda_S(t_1-\tau_1).
\label{eq:Born_Q_part_durations}
\end{align}

\section{Multitime non-Markovianity in the spin-boson model}
\label{sec:numerics}
We are now ready to apply the general expressions derived in Sec.~\ref{sec:analytic} to investigate deviations from the QRT in multitime probabilities within a paradigmatic open quantum system, namely the spin-boson model \cite{LeggettRMP1987}. This model provides a simple yet powerful framework to describe the interaction between a two-level system and a dissipative environment, capturing essential features of the evolution in diverse dynamical regimes.

\subsection{The model}
\label{subsec:spinboson}

We consider the spin--boson Hamiltonian
\begin{align}\label{eq:spinboson_H}
&H = H_S + H_B + H_{SB}\\
&= \frac{\omega_0}{2}\sigma_z + \sum_n \omega_n a_n^\dagger a_n
+ \sigma_x\otimes \sum_n \big(g_n a_n + g^*_n a_n^\dagger\big),\nonumber
\end{align}
i.e.\ a two-level system linearly coupled to a bosonic environment where $\sigma_i$, $i=x,y,z$, are the Pauli matrices
of the two-level system, $a_n, a^{\dagger}_n$ are the annihilation and creation operators of the $n$-th bosonic mode,
$\omega_0$ ($\omega_n$) is the free frequency of the system ($n$-th mode) and $g_n$ is the coupling between the two-level system and the $n$-th mode.
Assuming an initial global product state, see Eq.(\ref{eq:in}), and an initial thermal state of the environment 
at inverse temperature $\beta$
\begin{equation}\label{eq:thst}
   R= \bigotimes_n\frac{\exp\left[- \beta \, \omega_n a^{\dagger}_n a_n \right]}{Z_n},
\end{equation}
with 
\begin{equation}
    Z_n = \Tr\left[\exp\left[- \beta \omega_n a^{\dagger}_n a_n \right]\right],
\end{equation}
the reduced two-level system dynamics is fixed by the correlation function, see Eq.(\ref{eq:bath_corr_def}), 
\begin{align}
    C(t) =\, &\mathrm{Tr}_B\!\Big[\sum_{n,m}\big(g_n e^{-i \omega_n t} a_n + g_n^\ast e^{i \omega_n t} a_n^\dagger\big) \big(g_m a_m + g_m^\ast a_m^\dagger\big)\,R\Big]\nonumber\\
    = & \int_0^{+\infty} d\omega\  J(\omega) \left(e^{-i\omega t} (1+n_\beta(\omega))+ e^{i \omega t}
n_\beta(\omega)\right),\label{eq:cfsb} 
\end{align}
where we introduced the spectral density \cite{Breuer2002}
\begin{equation}\label{eq:jomega}
    J(\omega) = \sum_n |g_n|^2 \delta(\omega-\omega_n),
\end{equation}
and the mean occupation number at inverse temperature $\beta$ of the $n$-th mode,
\begin{equation}\label{eq:nbeta}
 n_\beta(\omega_n) := \frac{1}{e^{\beta \omega_n}-1} = \mathrm{Tr}_B\!\left[a^\dagger_n a_n R\right].
\end{equation}
The expression in Eq.(\ref{eq:cfsb}) allows one to readily take the continuum limit of the bosonic modes,
by simply replacing the spectral density defined in (\ref{eq:jomega}) with a positive continuous function of $\omega$, for $\omega \geq 0$ (and extending 
the definition in Eq.(\ref{eq:nbeta}) to a generic value of $\omega \geq 0$).

In particular, we take into account a Lorentzian spectral density centered at $\eta$ with width $\gamma$~\cite{BreuerKappler1999,Vacchini2010}:
\begin{equation}
J(\omega)=\frac{2\lambda^2\gamma}{(\omega-\eta)^2+\gamma^2},
\label{eq:J_lorentzian}
\end{equation}
which describes, for example, the coupling with a damped cavity mode. In fact, for such a choice of the spectral density,
the open-system dynamics can be equivalently obtained using an exact Lindbladian embedding in an enlarged
setting $S{+}M$ with a single damped pseudomode $M$~\cite{Garraway1997,Tamascelli2018,Albarelli2025}. Explicitly,
the reduced state $\rho_S(t)$ at time $t$ -- defined by Eq.(\ref{eq:Phi_def}) --
is equal to the reduced state
\begin{equation}
\rho_S(t)=\Tr_M[\rho_{SM}(t)],
\end{equation}
obtained from the $S+M$ state $\rho_{SM}(t)$ that is the solution of the master equation
\begin{align}
&\frac{d}{dt}\rho_{SM}(t)
= \mathcal L_{SM}[\rho_{SM}(t)]
:= -i\big[H_{\mathrm{SM}},\rho_{SM}(t)\big]\nonumber\\
&+ 2\gamma\,(n_{\beta}(\eta)+1)\,\mathcal D_b[\rho_{SM}(t)]+ 2\gamma\,n_{\beta}(\eta)\,\mathcal D_{b^\dagger}[\rho_{SM}(t)].
\label{eq:pseudomode_ME_thermal}
\end{align}
Here, the coherent system-pseudomode coupling is given by the Hamiltonian
\begin{equation}
H_{\mathrm{SM}}=\frac{\omega_0}{2}\sigma_z + \eta\,b^\dagger b + \lambda\,\sigma_x\big(b+b^\dagger\big),
\label{eq:H_PM}
\end{equation}
where $b$ and $b^\dagger$ are the annihilation and creation operators of the pseudomode,
while the dissipator is fixed by the operatorial structure
\begin{equation}
\mathcal D_L[\rho] := L\rho L^\dagger-\frac{1}{2}\{L^\dagger L,\rho\}.
\label{eq:lindblad_dissipator}
\end{equation}
Moreover, the initial condition is a product state, with a thermal state of the pseudomode, at inverse temperature $\beta$:
\begin{equation}
    \rho_{SM}(0) = \rho_S(0) \otimes \frac{\exp\left[- \beta \eta b^{\dagger} b \right]}
    {\Tr\left[\exp\left[- \beta \eta b^{\dagger} b \right]\right]};
\end{equation}
the case of a zero-temperature environment is indeed recovered for $\rho_M(0)=\ket{0}\bra{0}$
and $n_{\beta} = 0$.
The system-pseudomode evolution is thus fixed by a time-independent master equation, 
a Gorini-Kossakowski-Lindblad-Sudarshan master equation \cite{Gorini1976,Lindblad1976}, which means that the $S+M$ compound evolves in a Markovian (time-homogeneous) way; crucially, the reduced two-level system evolution will generally be non-Markovian: all the memory effects are encompassed in the interaction with the pseudomode. Even more, such a picture concerns not only the dynamics, but also the multitime expectation values -- being them correlators or sequential probabilities: Eq.(\ref{eq:C_A_def}) 
can be evaluated exactly by means of the QRT-expression on the enlarged $S+M$ system \cite{Chen2019,Smirne2022}, according to
\begin{align}
F_{\mathcal{A}_2,\mathcal{A}_1}(t_2,t_1)
&= \Tr_{SM}\Big[(\mathcal{A}_2\otimes\mathcal I_M)\circ
e^{(t_2-t_1)\mathcal{L}_{SM}}\,\nonumber\\
&\circ(\mathcal{A}_1\otimes\mathcal{I}_M)\,
\circ e^{t_1\mathcal{L}_{SM}}[\rho_{SM}(0)]\Big];\label{eq:twotqrt}
\end{align}
indeed, this expression includes all the violations of the QRT with respect to the reduced $S$ dynamical map.

Numerically, to solve the master equation (\ref{eq:pseudomode_ME_thermal}) and the two-time expression (\ref{eq:twotqrt}), 
we truncate the pseudomode Hilbert space to a finite Fock cutoff $n_{\max}$ and verify convergence of
all reported quantities upon increasing $n_{\max}$.
In particular, we work in the column-stacking (Liouville) representation and exploit QuTiP libraries \cite{Johansson2012,Johansson2013}.
The dynamics of the enlarged $S{+}M$ system is propagated through the exponential of the corresponding Lindblad generator, while multitime probabilities are computed by applying the relevant intervention maps on the system at the prescribed intermediate times, retaining the full $S{+}M$ state across the intervention, and propagating to the subsequent readout time under the same generator.

\subsection{Two-time non-Markovianity}
\label{subsec:pdiv_numerics}

The focus of our investigation is the characterization of the multitime probabilities emerging from sequential measurements
in general, non-Markovian, regimes where the QRT cannot be applied.
In particular, as explained in Sec.\ref{subsec:analytic_decomposition}, we use $\varepsilon_{\rm QRT}(t_2,t_1)$ 
defined in Eq.(\ref{eq:TV_def0}) to quantify the deviations
from the QRT and then the multitime non-Markovianity. 

\begin{figure}
  \centering
  \includegraphics[width=0.98\linewidth]{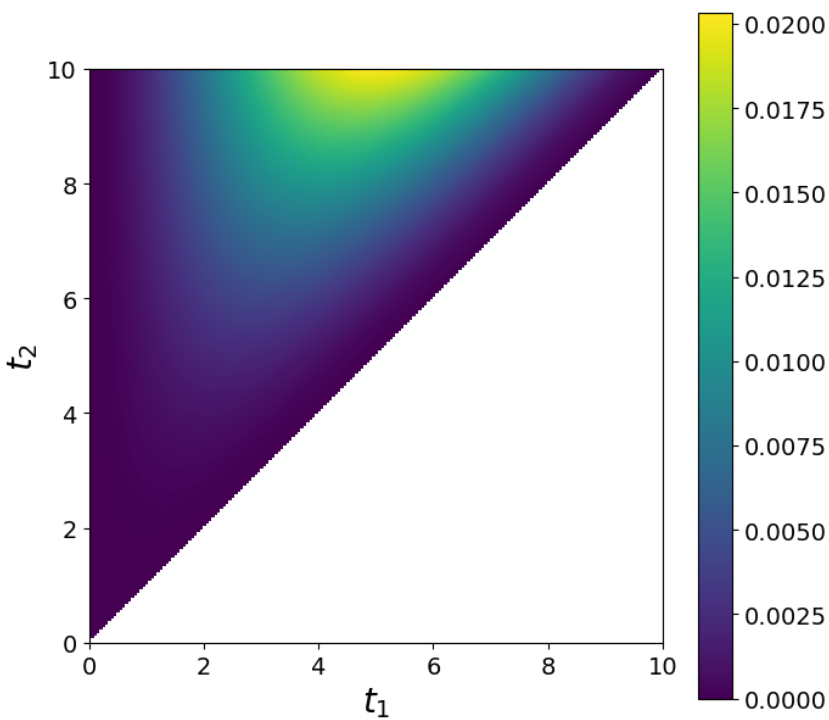}
  \caption{Two-time landscape of QRT violation.
  Heatmap of the QRT-violation quantifier
  $\varepsilon_{\rm QRT}(t_2,t_1)$ defined in Eq.(\ref{eq:TV_def0}
  over the $(t_1,t_2)$ plane, for $\gamma=0.1$, $\omega_0=\eta=4.5$,
  $\lambda=0.1$ [here and in the following all parameters are in arbitrary units],
  $\rho_S(0)=\ket{0}\bra{0}$ and a zero-temperature environment; the pseudomode levels are cut at $n_{max}= 8$.
  The white triangular region corresponds to $t_2<t_1$.}
  \label{fig:tv_t1t2_landscape}
\end{figure}

In Fig.~\ref{fig:tv_t1t2_landscape} we report $\varepsilon_{\rm QRT}(t_2,t_1)$ over the $(t_1,t_2)$ plane for a sequential projective measurement protocol in the $\sigma_z$ basis at both times. 
We observe that the QRT error is smallest close to the diagonal $t_2\simeq t_1$ and increases for
larger temporal separations, indicating that the impact of system--bath correlations across the intervention at
$t_1$ becomes more pronounced as the second readout is delayed. Moreover, the dependence on $t_1$ is nontrivial:
for fixed $t_2$ there is typically a window of intermediate $t_1$ values where $\varepsilon_{\rm QRT}$ is
maximized, reflecting the fact that the same intervention can be more or less disruptive depending on how much
correlation has built up between system and environment prior to $t_1$.

To better understand the non-Markovianity of the two-time statistics, we compare it with the non-Markovianity of the dynamics.
As stated in Sec.\ref{sec:sbd}, we identify the latter with a violation of the P-divisibility of the dynamics, i.e., 
the non-positivity of the propagator $\mathcal V(t_2,t_1)$ in Eq.(\ref{eq:divisibility_def}). Since we are dealing with a two-dimensional open quantum system,
we can take advantage of a direct characterization of the dynamical map and propagators associated with it.
In fact, any two-level quantum system state is univocally associated with a Bloch vector $\mathbf r\in\mathbb R^3$, $\|\mathbf r\| \leq 1$,
so that any map $\Lambda: \rho\mapsto \Lambda[\rho]$ is associated with an affine transformation \cite{Ruskai2002},
\begin{equation}
 \mathbf r \mapsto \mathbf r' = M\,\mathbf r + \mathbf v,
\label{eq:affine_num}
\end{equation}
where $M$ is a $3 \times 3$ real matrix and $\mathbf v \in\mathbb R^3$.
A two-level system map is trace preserving and positive if and only if it maps the Bloch ball into itself,
so that positivity of $\Lambda$ is equivalent to the requirement
\begin{equation}
\max_{\|\mathbf r\|\le 1}\,\|M\mathbf r+\mathbf v\|\le 1.
\label{eq:blochball_test_num}
\end{equation}
Thus, we quantify the non-positivity of the propagator $\mathcal V(t_2,t_1)$ as
\begin{equation}\label{eq:q_num}
q(t_2,t_1):=\max\!\left\{0,\ \max_{\mathbf r\in\mathcal S}\|M(t_2,t_1)\mathbf r+\mathbf v(t_2,t_1)\|-1\right\},
\end{equation}
where $M(t_2,t_1)$ and $\mathbf v(t_2,t_1)$ denote the affine representation of 
$\mathcal V(t_2,t_1)$ according to Eq.~(\ref{eq:affine_num}), and we estimate the maximum by sampling a set of directions
$\mathcal S=\{\mathbf r_\ell\}$ on the unit sphere using Fibonacci sampling \cite{Swinbank2006,Gonzalez2010}. 
In addition, to facilitate the comparison between two-time and one-time non-Markovianity, we consider the time-average quantifier of QRT violation, $\bar{\varepsilon}_{\rm QRT}(t_f)$ [see Eq.~(\ref{eq:epsQRTavg})], and introduce the corresponding average quantifier for the one-time case,
\begin{equation}
\bar{N}(\Lambda;t_f):=\frac{1}{t_f}\int_0^{t_f}dt_1\, q(t_f,t_1).
\label{eq:NPhi_num}
\end{equation}

\begin{figure*}
  \centering
  \includegraphics[width=0.32\linewidth]{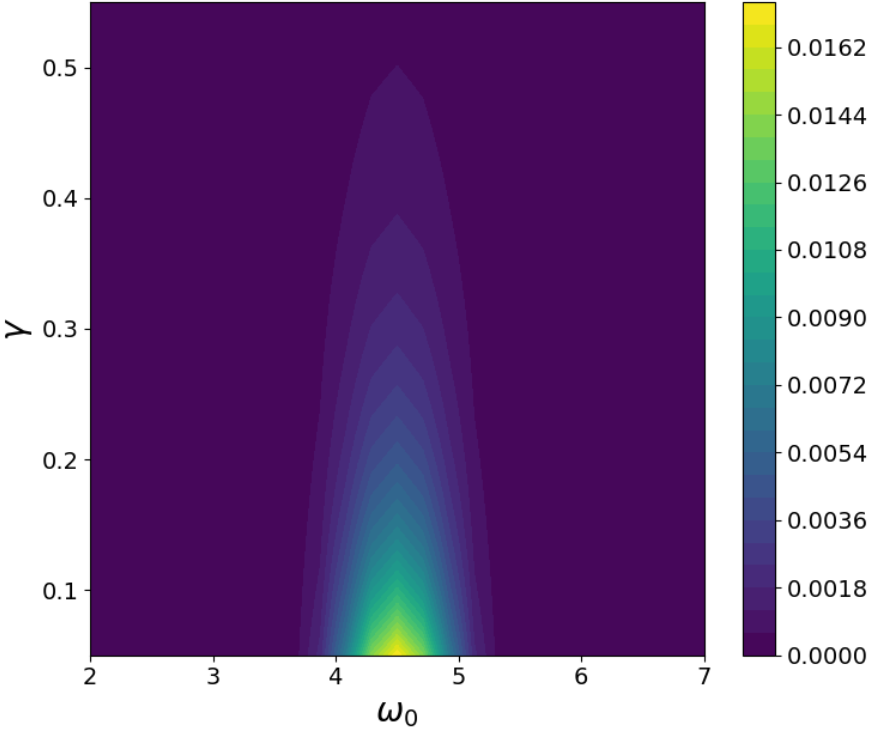}
  \includegraphics[width=0.32\linewidth]{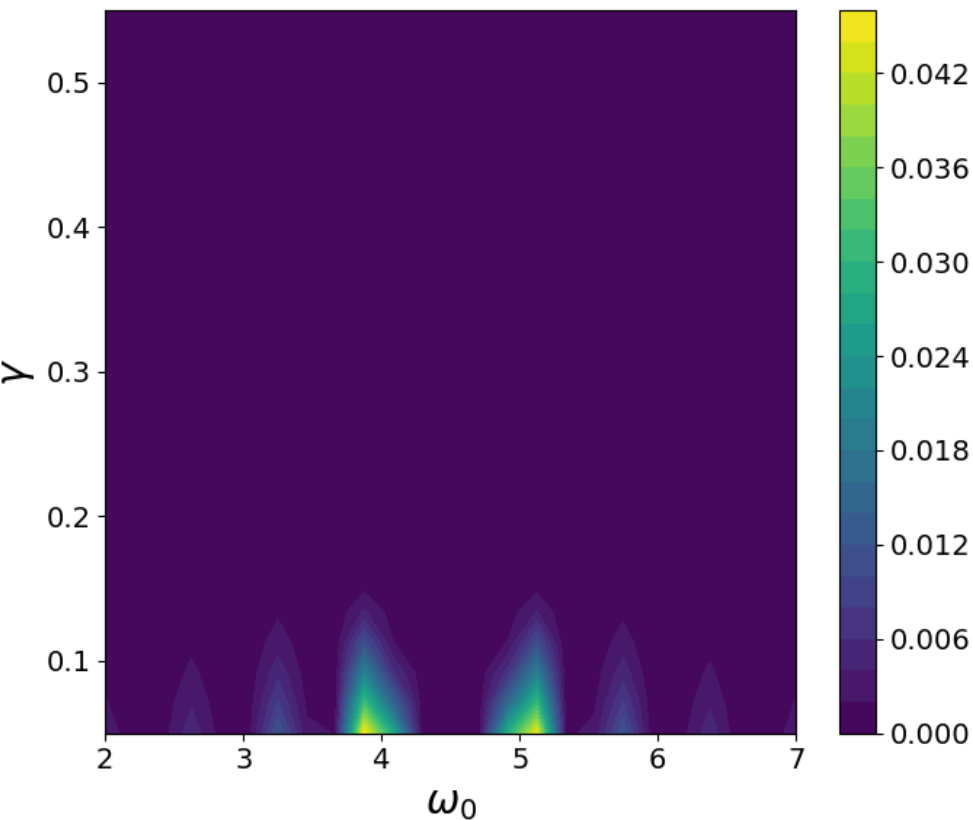}
      \includegraphics[width=0.32\linewidth]{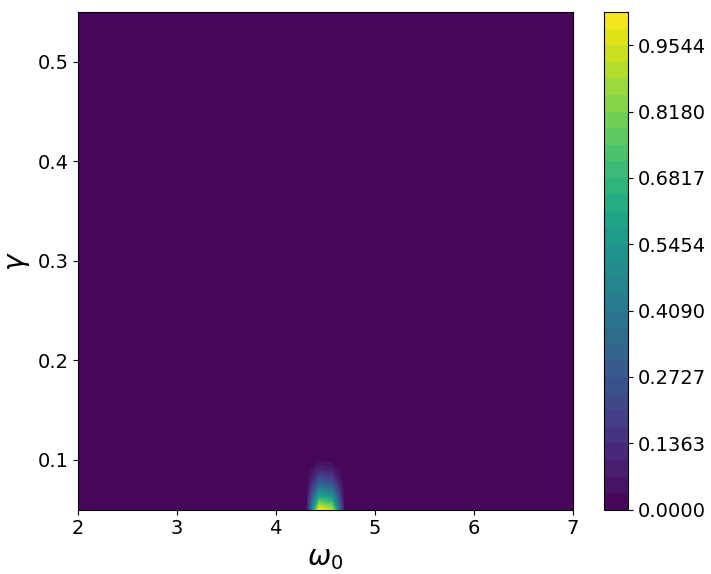}
  \caption{Average two-time non-Markovianity versus average one-time non-Markovianity.
  (Left) $\bar{\varepsilon}_{\rm QRT}(t_f)$ defined in Eq.(\ref{eq:epsQRTavg}) over the
  $(\omega_0,\gamma)$ plane, 
  for the sequential-measurement protocol with $\rho_S(0)=\ket{0}\!\bra{0}$ and projective $\sigma_z$ measurements at $t_1$ and $t_f=10$; $\eta = 4.5, \lambda=0.1, n_{max}=10$ and zero-temperature environment.
  (Central) $\bar{N}(\Lambda;t_f)$ defined in Eq.(\ref{eq:NPhi_num}) for the corresponding one-time reduced dynamics.
  (Right) Same quantity and values as the central panel, but for $t_f = 30$.
 }
  \label{fig:avg_vs_pindiv}
\end{figure*}

The left panel of Fig.~\ref{fig:avg_vs_pindiv} reports the average two-time non-Markovianity quantifier $\bar{\varepsilon}_{\rm QRT}(t_f)$ over the
$(\omega_0,\gamma)$ plane at fixed bath peak $\eta$ (and fixed $t_f,\lambda$, and measurement protocol). 
The plot highlights a region where QRT yields appreciable errors in sequential statistics.
As expected, the largest values occur for long bath memory (small $\gamma$), while in the short-memory regime
(large $\gamma$) the QRT error becomes negligible over the explored range of $\omega_0$.
The average one-time non-Markovianity, as quantified by $\bar{N}(\Lambda;t_f)$, 
is shown in the central and right panels of Fig.~\ref{fig:avg_vs_pindiv}. The central panel is characterized by the same
parameter values as the left panel, and the comparison between the two panels illustrates that
one-time and two-time non-Markovianity are related but not equivalent: while both quantifiers tend to increase
in long-memory regimes, their detailed dependence on $(\omega_0,\gamma)$ differs. Operationally,
$\bar{\varepsilon}_{\rm QRT}(t_f)$ probes multitime, protocol-dependent, memory across the intervention at
$t_1$, whereas $\bar{N}(\Lambda;t_f)$ is constructed solely from properties of the one-time map
$\Lambda_S(t)$.
The P-divisibility violation $\bar{N}(\Lambda;t_f)$ is confined to a narrow strip at small $\gamma$ and exhibits two off-resonant maxima, with a local minimum at $\omega_0=\eta$. By contrast, the QRT violation $\bar{\varepsilon}_{\rm QRT}(t_f)$ displays a single peak centered at resonance and extends to substantially larger values of $\gamma$. This shows that the two quantifiers probe different aspects of memory. The quantity $\bar{N}(\Lambda;t_f)$ is determined solely by the behavior of the reduced propagator and, in particular, by the loss of positivity of the intermediate map $\mathcal{V}(t_f,t_1)$,
indicating a back flow of information to the open system during the time interval $[t_1, t_f]$. 
The quantity $\bar{\varepsilon}_{\rm QRT}(t_f)$, instead, is sensitive to the system--environment correlations generated 
up to time $t_1$ and 
that subsequently impact the measured
sequential statistics. For the protocol considered here, this effect is strongest near resonance, where the system and pseudomode are most efficiently coupled, and thus stronger correlations are built up. Accordingly, there are parameter regions in which no appreciable deviation from $P$-divisibility is detected within the chosen observation window, while the sequential statistics still display a clear breakdown of the QRT factorization. This provides a direct illustration that operational multitime memory is not determined by one-time divisibility properties of the reduced dynamics alone.

The double-peak structure observed in $\bar{N}(\Lambda;t_f)$ at $t_f=10$ is consistent with a transient feature within the chosen observation window.
Near resonance, where the system and pseudomode are most efficiently coupled, the signature of an information backflow in the reduced propagator appears to build up on a comparatively longer timescale within the chosen observation window. At finite detuning, an earlier visible backflow contribution can instead emerge, giving rise to the two lateral maxima. For larger $t_f$, 
these transient differences become substantially less pronounced, indicating that the off-resonant enhancement is largely a feature of the chosen observation window. This trend is consistent with the right panel of Fig.~\ref{fig:avg_vs_pindiv}, where $\bar{N}(\Lambda;t_f)$ is reported for a larger value of $t_f$ and the double-peak structure is strongly reduced. By contrast, $\bar{\varepsilon}_{\rm QRT}(t_f)$ already exhibits a single resonance-centered peak at shorter $t_f$, suggesting that the breakdown of the QRT is governed by a shorter timescale, associated more directly with the buildup of system--environment correlations than with the completion of a full oscillation cycle
of the information exchange between the system and the environment.

\subsubsection*{Perturbative analysis: improvement from the second-order correction.}
The benchmark with the exact two-time statistics provided by the system plus pseudomode configuration further allows us
to check the quality of the second-order correction to the QRT we derived in Sec.\ref{sec:soe}.
In particular, we compare the exact two-time probability $p(a_2,t_2;\,a_1,t_1)$, evaluated 
via Eq.(\ref{eq:twotqrt}), and the approximated two-time probability
$p_{\lambda^2}(a_2,t_2;\,a_1,t_1)$, which includes the second-order correction 
to the QRT given by Eq.(\ref{eq:Born_Q_part_durations}), i.e., that is obtained by replacing the latter, along with 
Eqs.(\ref{eq:exact_PQ_split}), (\ref{eq:exact_QRT_like_part}) into the definition of the two-time probability, Eq.(\ref{eq:seq_prob_def_ij}), so that
\begin{align}
   p_{\lambda^2}(a_2,t_2;\,a_1,t_1) =  p_{\rm QRT}(a_2,t_2;\,a_1,t_1)+\nonumber\\
   \Tr_S\!\left[
\mathcal{M}^{(2)}_{a_2}\circ \tilde{\Phi}^{(Q)}_{\mathcal{M}^{(1)}_{a_1}}(t_2,t_1)[\rho_S(0)]
\right].
\end{align}
To quantify the difference between $p(a_2,t_2;\,a_1,t_1)$ and $p_{\lambda^2}(a_2,t_2;\,a_1,t_1)$, we use once again the Kolmogorov distance 
[compare with Eq.(\ref{eq:TV_def0})], namely,
\begin{equation}\label{eq:TV_defl2}
\varepsilon_{\lambda^2}(t_2,t_1)
:=
\frac{1}{2}\sum_{a_1,a_2}\big|\,p(a_2,t_2;\,a_1,t_1)-p_{\lambda^2}(a_2,t_2;\,a_1,t_1)\,\big|.
\end{equation}

\begin{figure}
  \centering
  \includegraphics[width=0.95\linewidth]{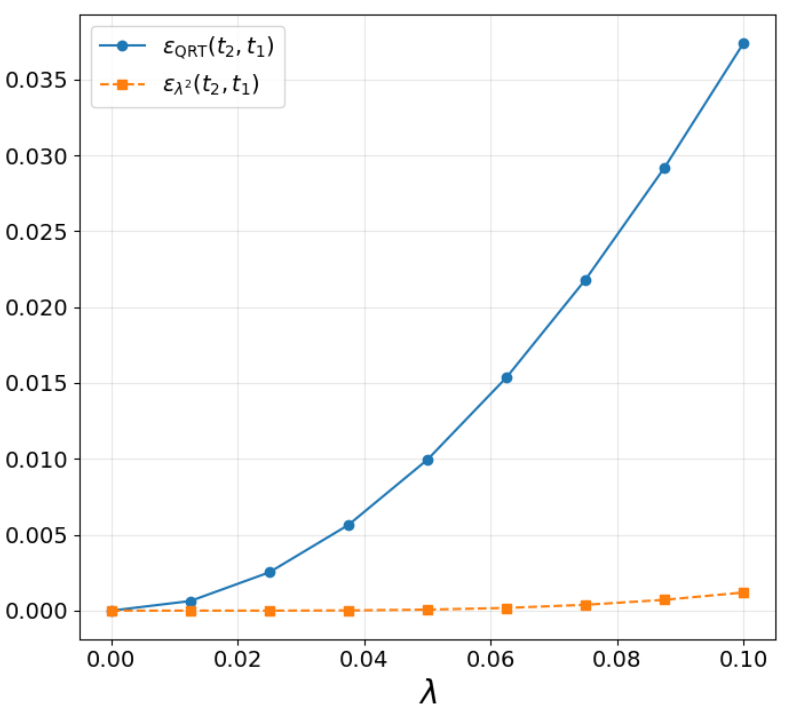}
  \caption{Second-order correction beyond-QRT.
QRT violation $\varepsilon_{\rm QRT}(t_2,t_1)$ as defined in Eq.(\ref{eq:TV_def0})
and residual error after including the second-order correction, $\varepsilon_{\lambda^2}(t_2,t_1)$ [Eq.(\ref{eq:TV_defl2})],
as a function of the coupling strength $\lambda$.
Parameters: $\gamma=0.1$, $\omega_0=\eta=4.5$, zero-temperature environment and 
$n_{max}=8$.
Protocol: $\rho_S(0)=\ket{+}\!\bra{+}$; Lüders projective measurement in the $\sigma_z$ basis at $t_1=5$ and readout in the $\sigma_x$ basis at $t_2=10$.}
  \label{fig:born_validation}
\end{figure}

Figure~\ref{fig:born_validation} shows that the second-order correction yields a
substantially smaller residual discrepancy, compared to the whole QRT violation $\varepsilon_{\rm QRT}(t_2,t_1)$, confirming that the leading memory term captured by
Eq.~\eqref{eq:Born_dressed_kernel_def} significantly moves the sequential statistics in the correct direction;
indeed, the residual error grows as the coupling strength increases, remaining anyway below $10^{-3}$ in the considered parameter regime.

\subsection{Protocol dependence: temperature, initial states and measurement bases}
\label{subsec:protocol_dependence}

We now analyze how the average QRT violation depends on the details of the multitime protocol, 
namely, the initial state, environmental temperature and the choice of the sequential measurements. 

First, we repeat the scan over the parameter space $(\omega_0,\gamma)$ at fixed final time $t_2=t_f$,
for different values of $\beta$;
moreover, we fix the protocol to sequential projective measurements in the $\sigma_z$ basis, with a
L\"uders instrument, and the initial state $\rho_S(0)=\ket{0}\!\bra{0}$.

\begin{figure*}
  \centering
  \includegraphics[width=0.95\linewidth]{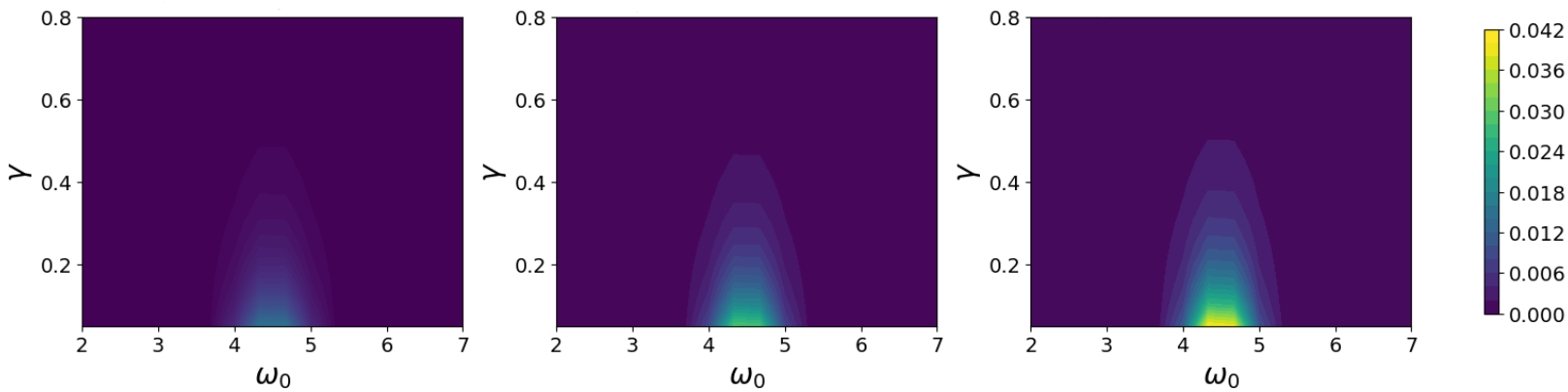}
  \caption{Temperature dependence of the average QRT violation.
 $\bar{\varepsilon}_{\rm QRT}(t_f)$ as defined in Eq.(\ref{eq:epsQRTavg})
  over the $(\omega_0,\gamma)$ plane at fixed bath peak $\eta=4.5$ and coupling $\lambda=0.1$,
  using the sequential measurement protocol
  consisting of $\rho_S(0)=\ket{0}\!\bra{0}$ and projective $\sigma_z$ measurements with L\"uders update. The three panels refer to different environmental temperatures:
  zero temperature (left panel), $\beta =0.25$ (central panel), corresponding to a mean occupation number
  $n_{\beta}=0.48$, and $\beta =0.154$ (right panel), corresponding to a mean occupation number
  $n_{\beta}=1.002$; the pseudomode has been correspondingly truncated at different values
  of the maximum level $n_{max}$: 8 (left), 13 (center) and 20 (right).
 }
  \label{fig:temperature_dependence}
\end{figure*}

Figure~\ref{fig:temperature_dependence} shows $\bar{\varepsilon}_{\rm QRT}$ for representative inverse
temperatures $\beta$, and hence thermal
occupations $n_{\beta}$. Across all cases, the region where QRT deviations are appreciable remains
localized in the long-memory regime (small $\gamma$) and is typically enhanced near resonance
$\omega_0\simeq\eta$. Increasing the temperature leads to a quantitative enhancement of the average QRT violation in the same
long-memory region, indicating that thermal fluctuations amplify the two-time memory 
for the present spin--boson setting. 
This behavior can be understood in terms of the 
$\mathcal P- \mathcal Q$ decomposition introduced in Sec. \ref{subsec:analytic_decomposition}.
At finite temperature the pseudomode is thermally populated, which increases the number of pathways through which
system--mode correlations can build up prior to intervention at time $t_1$.
As a consequence, the $\mathcal Q$-component of the joint state at $t_1$---the sole source of QRT violation in
Eq.~\eqref{eq:exact_memory_part}---can have a stronger impact on the observed subsequent statistics.

\begin{figure*}
    \centering
    \includegraphics[width=0.95\linewidth]{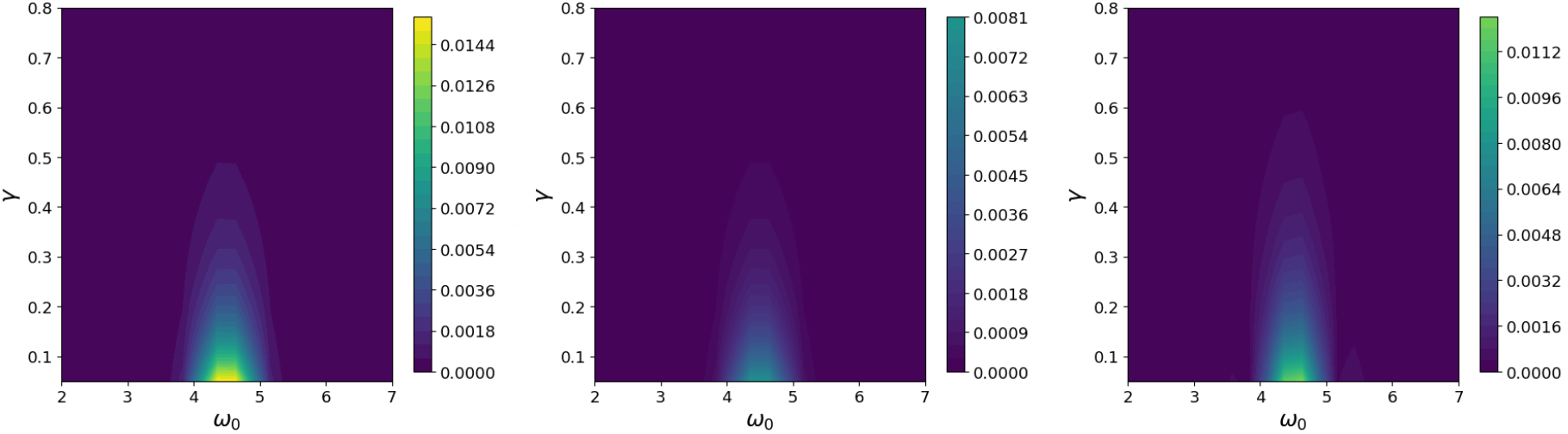}
    \caption{Protocol dependence of the average QRT violation.
  $\bar{\varepsilon}_{\rm QRT}(t_f)$ over the $(\omega_0,\gamma)$ plane for three sequential measurement protocols.
  Left: $\sigma_z$ basis, $\rho_S(0)=\ket{0}\!\bra{0}$ (reference protocol, same as Fig.~\ref{fig:avg_vs_pindiv} left).
  Center: $\sigma_z$ basis, $\rho_S(0)=\ket{+}\!\bra{+}$ (different initial state).
  Right: $\sigma_x$ basis, $\rho_S(0)=\ket{0}\!\bra{0}$ (different measurement basis).
  All other parameters are the same as in Fig.~\ref{fig:avg_vs_pindiv}.}
    \label{fig:protocol_dependence}
\end{figure*}

We next consider the dependence on the specific measurement protocol: The quantity 
$\bar{\varepsilon}_{\rm QRT}(t_f)$ is defined in terms of joint outcome probabilities and therefore depends on both the initial state and the measurement scheme. In particular, the initial preparation determines how correlations develop up to time $t_1$, 
while the choice of measurement basis affects both the back-action at $t_1$ and the components of the state probed at the final time.
To assess these effects, we repeat the parameter scan for two initial states, $\ket{0}$ and $\ket{+}=(\ket{0}+\ket{1})/\sqrt{2}$,
and for projective measurements in either the $\sigma_z$ or $\sigma_x$ basis (using the same basis at both times).
The results, shown in Fig.~\ref{fig:protocol_dependence}, indicate
that the magnitude of the average QRT error depends on the
protocol: changing either the initial state or the measurement basis can suppress or enhance
$\bar{\varepsilon}_{\rm QRT}(t_f)$, consistently with the fact that different protocols are sensitive to different
components of the memory due to the system-environment interaction. 
At the same time, the location of the region where QRT violations are largest
is comparatively robust: across all panels, it remains concentrated in the long-memory
regime (small $\gamma$) and near resonance $\omega_0\simeq \eta$.

The behavior in Fig.~\ref{fig:protocol_dependence} can be qualitatively understood from the relation between the measurement basis and the system–bath coupling.
In the present model, the interaction is proportional to $\sigma_x$, 
so that states aligned with the $\sigma_x$ basis experience reduced fluctuations of the coupling operator and tend to build weaker correlations with the environment before the intervention, which leads to smaller deviations from the QRT.
Conversely, preparations and measurements in a basis that does not commute with the coupling operator,
such as $\sigma_z$, generally enhances these fluctuations and make the sequential statistics more sensitive to system-environment correlations, resulting in larger QRT violations.

\subsection{Higher-order sequential-measurement memory: three-time QRT violations}
\label{subsec:three_time_check}
The two-time analysis conducted until now probes the QRT across a single intervention in the course of the evolution.
It is interesting, however, to look at the impact of a larger number of intermediate-time interventions,
which is what we are going to do now by
comparing the two-time QRT error with its three-time analogue.

Consider three interventions at $0\le t_1 \le t_2 \le t_3$ described by the instruments
$\{\mathcal{M}^{(j)}_{x_j}\}_{x_j}$ (CP, trace non-increasing maps -- see Sec.\ref{subsec:twotime_objects}). 
Generalizing Eq.~\eqref{eq:seq_prob_def_ij}, the exact joint distribution can be computed from the pseudomode embedding by propagating $\rho_{SM}$ and applying the corresponding instruments on $S$ at each intervention time:

\begin{align}
&p(x_3,t_3;\,x_2,t_2;\,x_1,t_1)= \Tr_{SM}\Big[(\mathcal{M}^{(3)}_{x_3}\otimes\mathcal I_M)\circ
e^{(t_3-t_2)\mathcal{L}_{SM}} \nonumber\\
&\,\circ(\mathcal{M}^{(2)}_{x_2}\otimes\mathcal I_M)\circ
e^{(t_2-t_1)\mathcal{L}_{SM}}\circ(\mathcal{M}^{(1)}_{x_1}\otimes\mathcal{I}_M)\,
\circ e^{t_1\mathcal{L}_{SM}}[\rho_{SM}(0)]\Big].
\end{align}
The QRT prediction generalizes indeed the two-time expression -- see Eqs.(\ref{eq:seq_prob_def_ij})
and (\ref{eq:QRT_like_factorization}) -- thus providing
\begin{align}
&p_{\rm QRT}(x_3,t_3;\,x_2,t_2;\,x_1,t_1)
=
\Tr_S\!\big[
\mathcal{M}^{(3)}_{x_3}\circ \Lambda_S(t_3-t_2)\nonumber\\
&\circ
\mathcal{M}^{(2)}_{x_2}\circ \Lambda_S(t_2-t_1)\circ\mathcal{M}^{(1)}_{x_1}\circ \Lambda_S(t_1)
[\rho_S(0)]
\big],
\label{eq:QRT_3time}
\end{align}
i.e.\ it concatenates the same one-time reduced map $\Lambda_S(\cdot)$ between successive interventions.
Also here, we quantify the three-time QRT error by the Kolmogorov distance
\begin{align}
\varepsilon^{(3)}_{\rm QRT}(t_3,t_2,t_1)
=&
\frac{1}{2}\sum_{x_1,x_2,x_3}
\big|
p(x_3,t_3;\,x_2,t_2;\,x_1,t_1) \nonumber\\
&-
p_{\rm QRT}(x_3,t_3;\,x_2,t_2;\,x_1,t_1)
\big|,
\label{eq:TV_3time}
\end{align}
and define the average version corresponding to uniform sampling [compare with Eq.(\ref{eq:epsQRTavg})],
\begin{equation}
\bar\varepsilon^{(3)}_{\rm QRT}(t_f)
:=
\frac{1}{t_f}\int_0^{t_f} dt_2\;
\frac{1}{t_2}\int_0^{t_2}\varepsilon^{(3)}_{\rm QRT}(t_f,t_2,t_1)\,dt_1;
\label{eq:avg3_fixed_tf}
\end{equation}
more explicitly, Eq.~\eqref{eq:avg3_fixed_tf} is the expected three-time QRT error for fixed $t_3 = t_f$, when $t_2$ is drawn uniformly in
$[0,t_f]$ and, conditional on $t_2$, the earlier time $t_1$ is drawn uniformly in $[0,t_2]$.

\begin{figure*}
  \centering
  \includegraphics[width=0.7\linewidth]{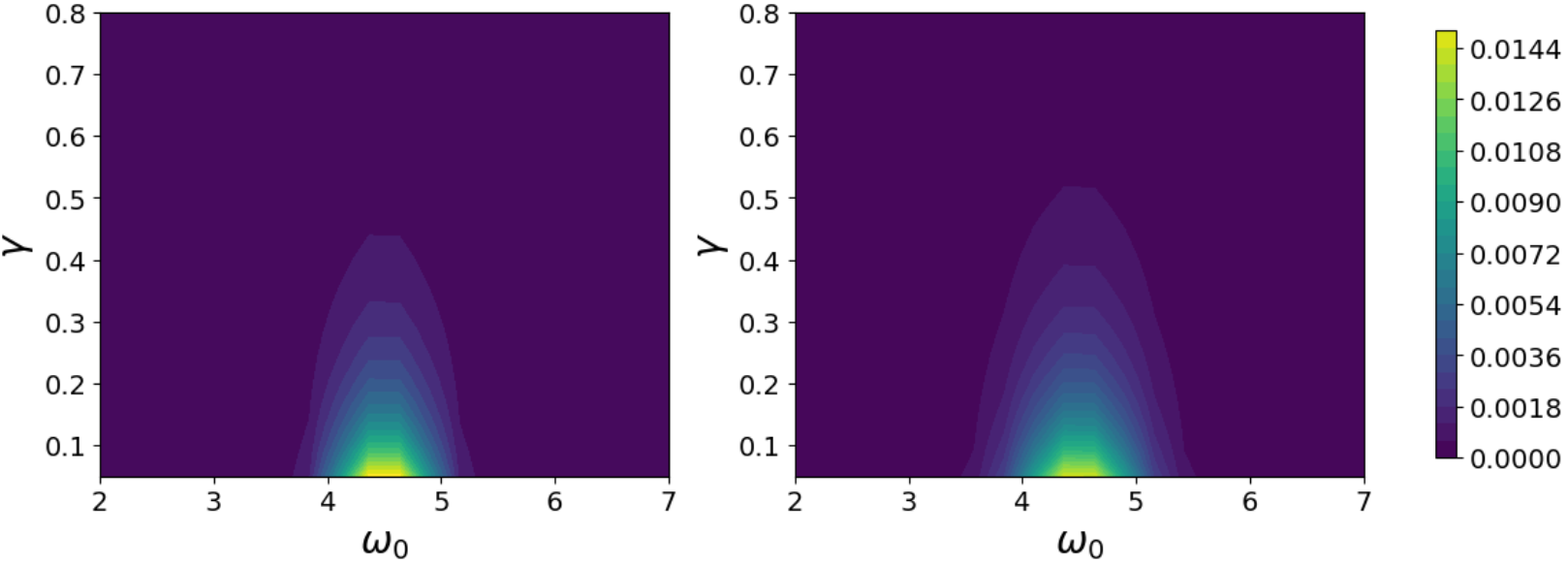}
    \includegraphics[width=0.7\linewidth]{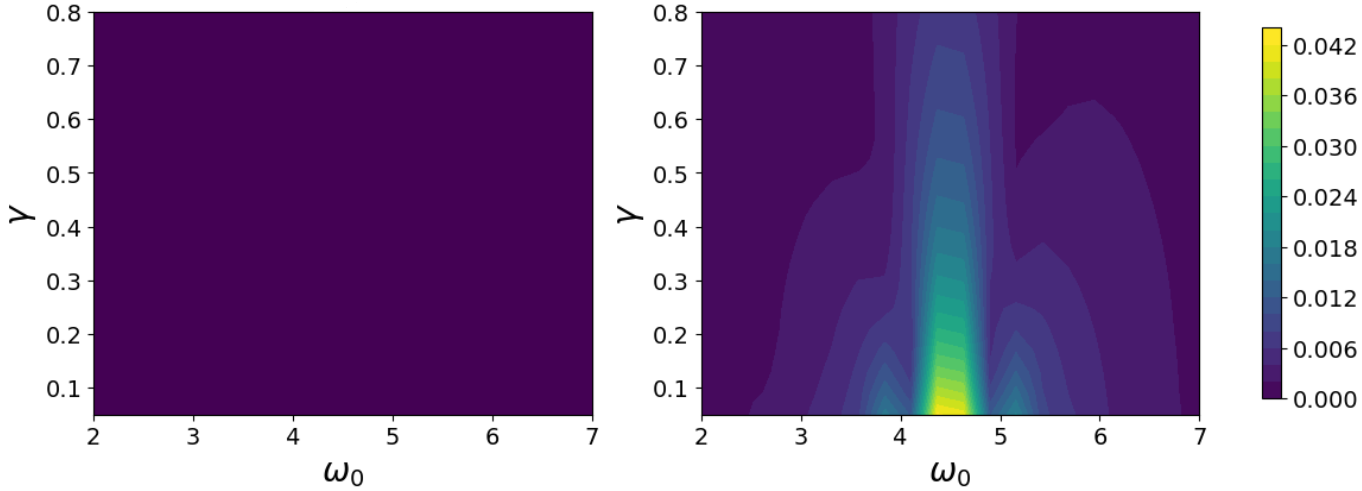}
  \caption{Two- vs three-time QRT error.
Left panels: two-time average QRT violation $\bar\varepsilon_{\rm QRT}(t_f)$
[Eq.~\eqref{eq:epsQRTavg}]. Right panels: three-time average QRT violation
$\bar\varepsilon^{(3)}_{\rm QRT}(t_f)$ [Eq.~\eqref{eq:avg3_fixed_tf}]. Both
quantities are shown over the $(\omega_0, \gamma)$ plane. The upper row corresponds to a commuting protocol with projective $\sigma_z$ measurements at all times. The lower row corresponds to a non-commuting protocol: $(\sigma_z,\sigma_x)$ in the two-time case (left) and $(\sigma_z,\sigma_x,\sigma_z)$ in the three-time case (right). Initial state $\rho_S(0) = \ket{0} \bra{0}$, zero-temperature environment, $\eta=4.5$, $\lambda=0.1$, $n_{max}=8$, $t_f=10.0$.}
  \label{fig:3vs2}
\end{figure*}

Figure~\ref{fig:3vs2} compares the average two-time error $\bar\varepsilon_{\rm QRT}(t_f)$ and the average three-time error $\bar\varepsilon^{(3)}_{\rm QRT}(t_f)$
for different measurement protocols: commuting (upper row) and non-commuting (lower row) projective measurements
at the intermediate times. In both cases,
the three-time error peaks in the same parameter region where the two-time error is maximal and
is overall larger, indicating that adding an intermediate intervention tends to make the QRT approximation worse,
especially where the environment retains memory.
Furthermore, by comparing the two different protocols we note that for non-commuting measurement protocols the two-time QRT error can be 
significantly 
reduced, while the corresponding three-time deviation is enhanced, 
with respect to the commuting measurements protocol. This represents a genuinely multitime manifestation of memory,
where the number of intermediate interventions does influence how the statistics depart from the Markovian description.
More specifically, in the non-commuting sequence the intermediate measurement in the $x$ basis constitutes a non-commuting intervention with respect to the $z$ preparation, the first measurement and the readout measurement.
The back-action of the $x$-basis measurement can modify the joint system--bath state in a way that enhances the contribution of the memory-carrying component (the $\mathcal Q$-part in the $\mathcal P$--$\mathcal Q$ decomposition) to the subsequent evolution.
The final measurement in the $z$ basis then acts as a readout that converts these correlations into observable differences in the three-time probabilities. This mechanism is not present when the 
$x$-basis measurement coincides with the final readout, since in that case the correlations do not propagate further into a subsequent measurement step.
In this sense, the choice of protocol determines which temporal order of the statistics is required to reveal non-Markovian features: a two-time probability may remain close to the QRT prediction, while
deviations become visible only at the level of three-time statistics.\\

\section{Conclusions and outlook}
\label{sec:conclusions}
We have analyzed multitime memory effects in open quantum systems through deviations from the QRT in sequential measurement statistics.
At the structural level, we showed that the two-time propagators and probabilities can be decomposed into a QRT-like contribution, fully
determined by the reduced dynamical map, and a complementary term that accounts for system--environment correlations across the intervention and thus the subsequent memory effects. 
This provides an explicit characterization of the part of the sequential dynamics that is not captured by one-time reduced evolution.

In the weak-coupling regime, we derived a second-order correction to the QRT, expressed in terms of the reduced map and bath correlation functions. In the spin--boson model, this correction significantly improves  the prediction of sequential statistics compared to the standard QRT approximation. Framing the analysis in terms of quantum instruments, we also introduced an operational quantifier of QRT breakdown based on the distance between exact and QRT-predicted joint probabilities.

The results highlight that multitime memory, as probed through sequential statistics, depends on the specific measurement protocol and is not fully characterized by one-time properties of the reduced dynamics, such as divisibility. In particular, different choices of initial state and measurement basis can lead to different sensitivities to system–environment correlations.
Moreover, higher-order sequential statistics can reveal memory effects that remain small at the level of a given two-time protocol.

Possible extensions of our analysis include the systematic investigation of higher-order temporal protocols, stronger-coupling corrections beyond the leading order term, and the treatment of initially correlated system--environment states. It would also be interesting to connect the present framework to broader notions of temporal nonclassicality formulated directly at the level of multitime probability distributions, for example through violations of Kolmogorov consistency conditions in sequential measurement protocols \cite{Milz2019,Strasberg2019,Milz2021,Strasberg2023,Luppi2026}.

\acknowledgements
This work has been supported by the Italian Ministry of Research and Next Generation
EU via the PRIN 2022 project Quantum Reservoir Computing (QuReCo)
(contract n. 2022FEXLYB), the PRIN 2022 Project EQWALITY (Contract N. 202224BTFZ),
and the NQSTI-Spoke1-BaC project QSynKrono (contract n. PE00000023-QuSynKrono).

\end{document}